\documentclass[aps,prl,reprint,showpacs]{revtex4-1}  
\usepackage{graphicx}  
\usepackage{dcolumn}   
\usepackage{bm}        
\usepackage{amssymb}   
\usepackage{draftcopy}

\begin{document}

\newcommand{\EuYSO}{Eu$^{3+}$:Y$_2$SiO$_5$~}
\newcommand{\Eu}{Eu$^{3+}$}
\newcommand{\YSO}{Y$_2$SiO$_5$}

\title{Absolute and relative stability of an optical frequency reference based on spectral hole burning in \EuYSO}

\author{David~R.~Leibrandt}
\email{david.leibrandt@nist.gov}

\author{Michael~J.~Thorpe}
\thanks{Present address: Bridger Photonics, 2310 University Way, Bozeman, MT 59715, USA}

\author{Chin-Wen~Chou}
\author{Tara~M.~Fortier}
\author{Scott~A.~Diddams}
\author{Till~Rosenband}
\affiliation{National Institute of Standards and Technology, 325 Broadway St., Boulder, CO 80305, USA}

\date{\today}

\begin{abstract}
We present and analyze four frequency measurements designed to characterize the performance of an optical frequency reference based on spectral hole burning in \EuYSO. The first frequency comparison, between a single unperturbed spectral hole and a hydrogen maser, demonstrates a fractional frequency drift rate of $5 \times 10^{-18}$ s$^{-1}$. Optical-frequency comparisons between a pattern of spectral holes, a Fabry-P\'erot cavity, and an Al$^+$ optical atomic clock show a short-term fractional frequency stability of $1 \times10^{-15} \tau^{-1/2}$ that averages down to $2.5^{+1.1}_{-0.5} \times 10^{-16}$ at $\tau = 540~s$ (with linear frequency drift removed). Finally, spectral hole patterns in two different \EuYSO crystals located in the same cryogenic vessel are compared, yielding a short-term stability of $7 \times10^{-16} \tau^{-1/2}$ that averages down to $5.5^{+1.8}_{-0.9} \times 10^{-17}$ at $\tau = 204$~s (with quadratic frequency drift removed).

\end{abstract}

\maketitle


Frequency-stable laser local oscillators (LLOs) are important tools for a variety of precision measurements.  Examples include both scientific applications such as searches for the variation of fundamental constants \cite{Blatt2008,Rosenband2008} and tests of general relativity \cite{Hough2005,Reynaud2009} as well as technical applications such as synthesis of low-phase-noise microwaves \cite{Zhang2010,Fortier2011,Benedick2012} and relativistic geodesy \cite{Chou2010b,Bondarescu2012}.  The stability of the best lasers constructed to date \cite{Jiang2011,Kessler2012,Nicholson2012} is limited by thermomechanical length fluctuations of the Fabry-P\'{e}rot reference cavities used for stabilization (i.e., thermally-driven displacement fluctuations of the atoms that make up the cavity) \cite{Numata2004,Kessler2012a}.  Present and future applications would benefit from lasers with improved stability \cite{Nicholson2012,Hinkley2013}.

One alternative to the mechanical frequency reference provided by Fabry-P\'{e}rot cavities is spectral-hole burning (SHB) laser-frequency stabilization \cite{Sellin1999a,Strickland2000,Pryde2002,Bottger2003,Julsgaard2007}.  In this technique, the frequency reference is an atomic transition of rare-earth dopant ions in a cryogenically cooled crystal.  These systems typically display an inhomogeneously broadened absorption line with a linewidth of order gigahertz, but the optical coherence time can be as long as milliseconds.  Thus, in systems with an appropriate level structure, a spectrally-narrow transparency, or spectral hole, can be written into the broad absorption line by optically pumping a subset of the dopant ions to a long-lived auxiliary state.  For laser-frequency stabilization, a spectral hole is written into the crystal at the beginning of the experiment, and subsequently the laser frequency is stabilized to the center of the spectral hole by probing the transmission of the crystal and feeding back to the laser frequency.  Because the coupling between crystal strain and the internal states of the dopant ions is weak, the thermomechanical noise that limits Fabry-P\'{e}rot cavities is only a weak perturbation to the frequencies of spectral holes.

The material system \EuYSO in particular is very promising for high-performance laser-frequency stabilization because it supports spectral holes with a linewidth as narrow as 122~Hz \cite{Equall1994} and a lifetime of order $10^6$~s at 4~K \cite{Konz2003}.  In addition, high-resolution spectroscopy has shown that the environmental sensitivity of 580~nm spectral holes in \EuYSO to temperature, pressure, and accelerations are all smaller than that of Fabry-P\'{e}rot cavities \cite{Thorpe2011,Chen2011}.  Furthermore, the sensitivity to magnetic-field fluctuations and perturbations to the spectral hole pattern due to side-holes and anti-holes are small enough to allow laser frequency stabilization at the $10^{-17}$ fractional frequency level \cite{Thorpe2013}.

\begin{figure}
\includegraphics[width=1.0\columnwidth]{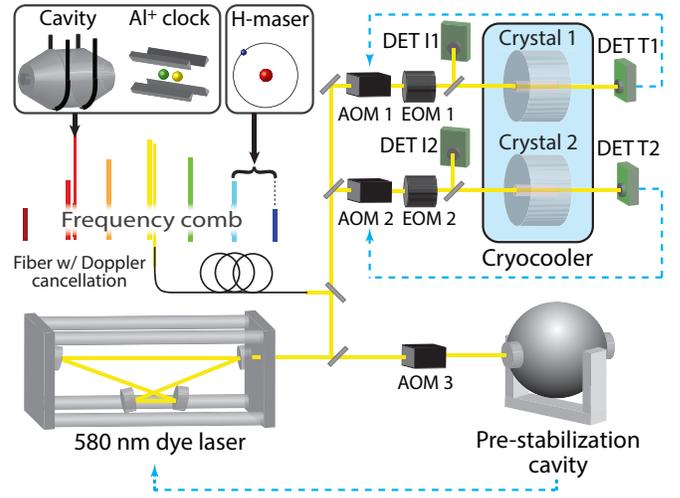}
\caption{\label{fig1} Experimental setup for laser frequency stabilization to a pattern of \EuYSO spectral holes and comparisons with other frequency references.  Solid yellow lines denote laser propagation and blue dashed lines denote frequency feedback paths.  AOM: acousto-optic modulator, EOM: electro-optic modulator, DET: detector.}
\end{figure}

\begin{figure*}
\includegraphics[width=1.0\textwidth]{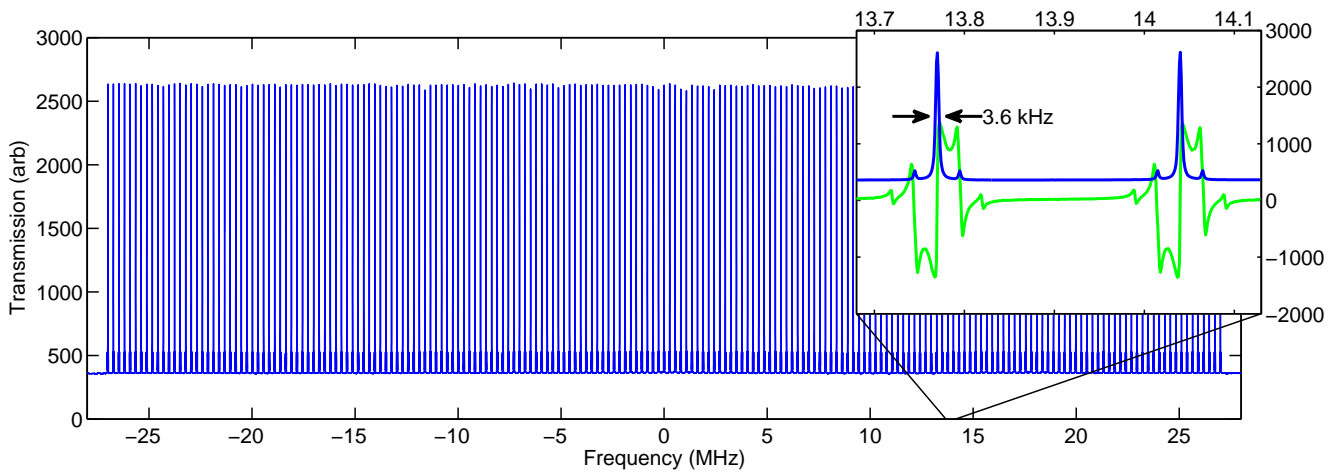}
\caption{\label{Fig9} A pattern of 201 \EuYSO spectral holes used for laser-frequency stabilization.  The inset shows the transmission (blue) and PDH error signal (green) of two holes selected from the pattern.  This measurement is performed after stabilization of the laser to the spectral hole pattern for approximately 110 minutes, at which time the spectral holes have been broadened by the probe light from an initial FWHM of 2.0~kHz to a FWHM of 3.6~kHz.}
\end{figure*}

The best SHB LLO performance demonstrated previously \cite{Thorpe2011} employed a two-stage laser frequency stabilization scheme where the laser is first pre-stabilized to a Fabry-P\'{e}rot cavity, then the pre-stabilized laser is used to burn and probe spectral holes in \EuYSO.  Pre-stabilization enabled the observation of spectral holes as narrow as 500~Hz, corresponding to a quality factor of $10^{12}$.  In order to slow the broadening of the spectral holes due to repeated probing the laser was stabilized to a pattern of spectral holes in a pulsed manner, where the laser spent 1~ms probing each side of a randomly selected spectral hole.  A comparison measurement of this SHB LLO with an independent reference laser demonstrated a stability of $6 \times 10^{-16}$ for measurement durations between 2~s and 8~s.  The stability of this measurement was limited, however, by the frequency noise of the cavity-stabilized reference laser used for the comparison, so the actual frequency stability of the SHB LLO at long averaging times remained unknown.

Here, we extend these results to include frequency-comparison measurements with a hydrogen maser, an aluminum-ion optical atomic clock, and a second SHB LLO.  Each of these references provides information about the frequency stability of the SHB LLO at different averaging times and with different measurement sensitivities.  The cavity-stabilized reference laser \cite{Young1999} has the best stability at averaging times below approximately 10~s.  The Al$^+$ clock \cite{Chou2010a} has the best stability at averaging times above 10~s, but it is limited to averaging times below 1000~s by ion loss.  Finally, the hydrogen maser \cite{Parker1999b,Parker2005} has a stability worse than the Al$^+$ clock, but it is very reliable and it can be used to measure the stability of the SHB LLO at arbitrarily long averaging times.  In the comparison between two SHB LLOs, much of the environmental noise is common mode but thermomechanical noise in the crystals is independent, thus the measurement provides an upper bound on the thermomechanical noise floor of SHB laser-frequency stabilization.  Together, these measurements provide a comprehensive characterization of the SHB LLO frequency noise down to the $10^{-17}$ fractional frequency level.

In addition, we have improved two aspects of the experimental setup: the \EuYSO crystal is housed in a closed-cycle cryostat with improved temperature stability and low levels of vibration \cite{Thorpe2013}, and we use the Pound-Drever-Hall (PDH) method for laser frequency stabilization \cite{Drever1983,Julsgaard2007}.  The PDH method is advantageous over the side-lock method used previously \cite{Thorpe2011} because it achieves a higher servo bandwidth and is less sensitive to low-frequency laser intensity noise, thus allowing a tighter lock of the laser to the \EuYSO spectral-hole pattern.


The experimental setup for this work has been described previously \cite{Thorpe2013}, so here we present only a brief summary (see Fig.~\ref{fig1}).  We burn and probe site 1 spectral holes \cite{Konz2003} in two crystals of 1.0~atomic~(at.)~\%~\EuYSO located in the same 3.65~K cryogenic vessel with a single 580~nm dye laser.  The dye laser is pre-stabilized to a compact Fabry-P\'{e}rot reference cavity \cite{Leibrandt2011a} with a fractional frequency stability of $1.2 \times 10^{-15}$ between 0.5 and 12~s and a typical drift rate of $10^{-14}$/s.  The frequency and intensity of the laser incident on each crystal can be tuned independently by manipulating the radio-frequency (RF) drives of AOMs~1 and 2.  Phase-modulation sidebands are imprinted on the lasers for PDH laser frequency stabilization by electro-optic modulators (EOMs)~1 and 2.  Detectors (DETs)~I1 and I2 measure the power incident on the crystals, which is stabilized at the desired level by feedback to the amplitudes of the RF drives of AOM~1 and AOM~2.  Finally, detectors~T1 and T2 measure the transmission of the laser through crystal~1 and crystal~2, respectively.

The error signal for laser frequency stabilization to a spectral-hole pattern is derived by the PDH method \cite{Drever1983,Julsgaard2007} in transmission by use of phase modulation at 25~kHz with a modulation depth of 0.12, such that each of the first-order phase modulation sidebands have 6~\% of the optical power contained in the carrier.  In order to slow degradation of the spectral hole pattern by the laser used to probe the spectral holes, for laser frequency stabilization we use a pattern of 201 holes spaced by 270~kHz (see Fig.~\ref{Fig9}).  The laser probes a randomly selected hole for 1~ms, then frequency feedback is applied by shifting the drive frequencies of  AOM~1 and AOM~2, and finally a new spectral hole is selected.  The duty cycle of the probe laser is 85~\%.  With a probe intensity of 2.0~$\mu$W cm$^{-2}$, this allows us to run the frequency servo for 2 hours before spectral hole broadening by the probe laser starts to degrade the performance of the spectral hole pattern as a frequency reference.


\begin{figure}
\includegraphics[width=1.0\columnwidth]{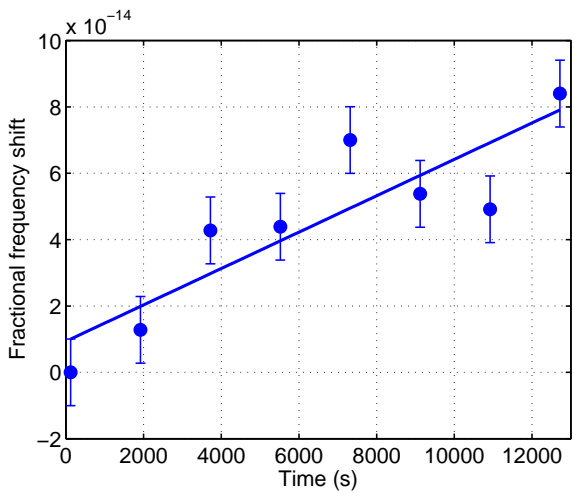}
\caption{\label{fig7} Frequency comparison of a single unperturbed \EuYSO spectral hole with a hydrogen maser (blue circles).  This measurement demonstrates that when left unperturbed, spectral holes can have very low frequency drift rates; the blue line is a linear fit with a slope of $5 \times 10^{-18}$~s$^{-1}$.  The scatter of the data is due primarily to short time frequency noise of the hydrogen maser.}
\end{figure}

In order to determine the frequency stability of spectral holes in the absence of probing, we burn a single spectral hole and measure its frequency once every 30~minutes relative to a hydrogen maser \cite{Parker1999b,Parker2005}.  The results are shown in Fig.~\ref{fig7}.  The scatter of the data is due primarily to short-term frequency noise of the hydrogen maser.  However, the long-term maser fractional frequency drift rate is less than $1 \times 10^{-20}$~s$^{-1}$.  Therefore the measured drift rate of $5 \times 10^{-18}$~s$^{-1}$ is primarily due to the spectral hole.  This drift rate is roughly an order of magnitude smaller than that of a typical room-temperature Fabry-P\'{e}rot cavity, and is similar to that reported by \citet{Chen2011} for spectral holes in \EuYSO.

\begin{figure}
\includegraphics[width=1.1\columnwidth]{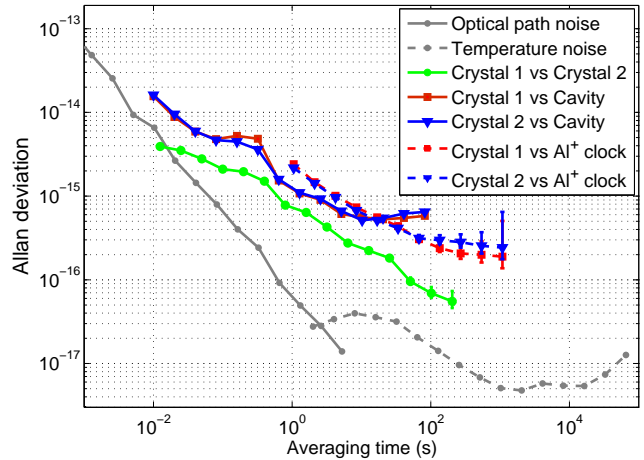}
\caption{\label{fig8} Optical frequency comparisons of a laser locked to \EuYSO spectral hole patterns with several independent optical frequency references.  A comparison with a second laser locked to a Fabry-P\'{e}rot cavity (solid red and blue lines, 3400~s of data with linear drift removed) shows the absolute performance of the spectral hole reference at short times; at longer times the measurement sensitivity is limited at $6 \times 10^{-16}$ by the thermomechanical noise floor of the cavity.  A comparison with an Al$^+$ optical clock (dotted red and blue lines) indicates the absolute performance of the spectral hole reference at long times, but the short-term measurement sensitivity is limited by the stability of the Al$^+$ clock and the measurement duration is limited by the lifetime of the Al$^+$ ion in the trap.  Finally, a comparison between two laser beamlines locked independently to spectral hole patterns in two crystals located in the same cryostat (green line, 1900~s of data with quadratic drift removed), where much of the technical noise is common mode, indicates the performance of the spectral hole frequency servo at short times and an upper bound of the spectral hole thermomechanical noise floor at long times.  The predicted frequency noise due to laser path length fluctuations (i.e., Doppler shifts due to vibrations of the mirror mounts and cryostat) and cryostat temperature fluctuations (converted into frequency noise using the temperature sensitivity measured in Ref.~\cite{Thorpe2013}) is shown in gray.}
\end{figure}

Figure~\ref{fig8} shows measurements of the frequency noise of a laser locked to a pattern of spectral holes in \EuYSO relative to a Fabry-P\'{e}rot cavity \cite{Young1999} and an Al$^+$ optical atomic clock \cite{Chou2010a}.  The comparison with the Fabry-P\'{e}rot cavity shows a short-term stability of $1 \times10^{-15} \tau^{-1/2}$ for $0.01~\textrm{s} < \tau < 10~\textrm{s}$.  At longer durations, this measurement is limited by the thermomechanical noise floor of the cavity at a level of $6 \times 10^{-16}$.  The comparison with the Al$^+$ optical clock is limited in stability by the Al$^+$ clock but averages down to $2.5^{+1.1}_{-0.5} \times 10^{-16}$ at $\tau = 540~s$.  The duration of this measurement was limited by ion loss in the Al$^+$ optical clock.  These measurements represent an upper bound on the absolute stability of the SHB LLO, after subtracting a linear frequency drift of $1.9 \times 10^{-17}$~s$^{-1}$ for crystal 1 and $2.7 \times 10^{-17}$~s$^{-1}$ for crystal 2.  The linear frequency drift of the laser locked to the pattern of spectral holes seen here is larger than the linear frequency drift of an unperturbed spectral hole seen in the comparison with a hydrogen maser (Fig.~\ref{fig7}) due to a small bias in the frequency-servo error signal, which pushes the center frequency of the spectral holes \cite{Julsgaard2007}.  Note that the errorbars on the Allan deviation are generated using $\chi^2$ statistics with one degree of freedom removed to account for the linear frequency drift removal \cite{Riley2008}.  We expect that the Allan deviation of this SHB LLO compared with a hypothetical perfect frequency reference would roughly follow the crystal versus cavity curves for averaging times from 10~ms to 10~s and the crystal versus Al$^+$ clock curves from 100~s to 1000~s.

We obtain information about the SHB frequency servo performance and an upper limit on the thermomechanical noise floor of this system by comparing the frequency of two laser beamlines that are independently locked to spectral hole patterns in the two crystals.  This measurement is insensitive to many common-mode environmental frequency shifts, including those due to vibrations and temperature fluctuations of the cryostat.  However, noise due to the frequency servos (i.e., photon shot noise and tracking errors) and thermomechanical noise are independent for the two SHB crystals.  In Fig.~\ref{fig8}, we plot the Allan deviation of the frequency difference between the two beamlines divided by $\sqrt{2}$, which is an upper bound for the frequency instability of the better beam.  Quadratic frequency drift is removed before calculation of the Allan deviation because, despite increasing the statistical uncertainty, this improves the upper bound on the thermomechanical noise.  Again the errorbars are generated using $\chi^2$ statistics, here removing two degrees of freedom to account for the quadratic frequency drift removal.  Note that while drift removal is known to introduce bias in the Allan deviation, this is likely at the 10~\% level for the maximum ratio of averaging time to measurement duration considered here \cite{Greenhall1997}.  In this measurement we see a short-term stability of $7 \times10^{-16} \tau^{-1/2}$ that averages down to $5.5^{+1.8}_{-0.9} \times 10^{-17}$ at 204~s.  At short averaging time the improvement of this measurement over the absolute frequency noise measurements may be due to Doppler shifts caused by vibrations of the cryostat that are common to the two crystals.  The measured frequency noise at 204~s indicates that the thermomechanical noise floor of this SHB LLO is below $7.3 \times 10^{-17}$ with 84~\% confidence, which is better than the frequency stability of the best reported Fabry-P\'{e}rot cavities \cite{Jiang2011,Kessler2012,Nicholson2012}.


In summary, we have demonstrated SHB laser frequency stabilization with a stability of $1 \times10^{-15} \tau^{-1/2}$ and an absolute fractional frequency noise of $2.5^{+1.1}_{-0.5} \times 10^{-16}$ at $\tau = 540~s$ (with linear frequency drift removed).   Furthermore, the thermomechanical noise floor of \EuYSO spectral holes has been shown to be below $7.3 \times 10^{-17}$ (with 84~\% confidence) and we are not aware of any fundamental obstacles to reaching such a laser frequency stability using SHB in \EuYSO \cite{Thorpe2013}.  This would represent a step beyond the present-day state-of-the-art in laser frequency stabilization, although there are promising new designs for Fabry-P\'erot cavities with reduced thermomechanical noise under development \cite{Kessler2012,Cole2013}.  Future work will include an optimization of the SHB laser frequency servo and the use of dense, steady-state spectral hole patterns in an effort to achieve high-stability, low-drift LLOs for optical atomic clocks.


We thank J.~Bergquist, S.~Brewer, S.~Cook, and D.~Wineland for useful discussions.  This work is supported by the Defense Advanced Research Projects Agency and the Office of Naval Research, and is not subject to US copyright.


%

\end{document}